\documentclass[]{hdsr}

\graphicspath{{figs/}}

\begin{document}

% Larger bottom margin for the first page
\newgeometry{bottom=1.5in}

% Editorial staff will replace the following values:
% 1. Volume number
% 2. Issue number
% 3. Article DOI
% e.g. for Volume 2, Issue 3, DOI 12.345:
% \volumeheader{2}{3}{12.345}
\volumeheader{Special}{6}{10.1162/99608f92.db29c137}

\begin{center}

  \title{Toward a Principled Framework for Disclosure Avoidance}
  \maketitle

  % Start page numbering on second page. Must appear *after* \maketitle
  \thispagestyle{empty}
  
  \vspace*{.2in}

  % Authors and Affiliations
  \begin{tabular}{cc}
    Michael B Hawes\upstairs{\affilone,*}, 
    Evan M Brassell\upstairs{\affilone}, 
Anthony Caruso\upstairs{\affilone},
Ryan Cumings-Menon\upstairs{\affilone},\\
Jason Devine\upstairs{\affilone}, 
Cassandra Dorius\upstairs{\affilone,\affiltwo},
David Evans\upstairs{\affilone,\affilthree},
Kenneth Haase\upstairs{\affilone}, \\
Michele C Hedrick\upstairs{\affilone},
Scott H Holan\upstairs{\affilone,\affilfour}, 
Cynthia D Hollingsworth\upstairs{\affilone},
Eric B Jensen\upstairs{\affilone}, \\
Dan Kifer\upstairs{\affilone,\affilfive},
Alexandra Krause\upstairs{\affilone},
Philip Leclerc\upstairs{\affilone}, 
James Livsey\upstairs{\affilone},\\
Rolando A Rodríguez\upstairs{\affilone},
Luke T Rogers\upstairs{\affilone},
Matthew Spence\upstairs{\affilone},
Victoria Velkoff\upstairs{\affilone},\\
Michael Walsh\upstairs{\affilone},
James Whitehorne\upstairs{\affilone},
and Sallie Ann Keller\upstairs{\affilone,\affilthree}
   \\[0.25ex]
   {\small \upstairs{\affilone} U.S. Census Bureau} \\
   {\small \upstairs{\affiltwo} Iowa State University} \\
   {\small \upstairs{\affilthree} University of Virginia} \\
    {\small \upstairs{\affilfour} University of Missouri} \\
     {\small \upstairs{\affilfive} Penn State University} \\
  \end{tabular}
  
  % Replace with corresponding author email address
  \emails{
    \upstairs{*}michael.b.hawes@census.gov 
    }
  \vspace*{0.4in}

\begin{abstract}
Responsible disclosure limitation is an iterative exercise in risk assessment and mitigation. From time to time, as disclosure risks grow and evolve and as data users’ needs change, agencies must consider redesigning the disclosure avoidance system(s) they use. Discussions about candidate systems often conflate inherent features of those systems with implementation decisions independent of those systems. For example, a system’s ability to calibrate the strength of protection to suit the underlying disclosure risk of the data (e.g., by varying suppression thresholds) is a worthwhile feature regardless of the independent decision about how much protection is actually necessary. Having a principled discussion of candidate disclosure avoidance systems requires a framework for distinguishing these inherent features of the systems from the implementation decisions that need to be made independent of the system selected. For statistical agencies, this framework must also reflect the applied nature of these systems, acknowledging that candidate systems need to be adaptable to requirements stemming from the legal, scientific, resource, and stakeholder environments within which they would be operating. This article proposes such a framework. No approach will be perfectly adaptable to every potential system requirement. Because the selection of some methodologies over others may constrain the resulting systems' efficiency and flexibility to adapt to particular statistical product specifications, data user needs, or disclosure risks, agencies may approach these choices in an iterative fashion, adapting system requirements, product specifications, and implementation parameters as necessary to ensure the resulting quality of the statistical product.
\end{abstract}
\end{center}

\vspace*{0.15in}
\hspace{10pt}

\section*{Media Summary}
Statistical agencies use a variety of statistical techniques to protect the confidentiality of the information entrusted to them by the individuals, businesses, and other entities that respond to their censuses and surveys. These techniques are \textit{disclosure avoidance} methods designed to make it hard to isolate the information about particular individuals or businesses in the statistical products that the agencies publish. The application of these methods, however, can have notable implications for the resulting accuracy and availability of the statistics to be released. 

There are a wide variety of disclosure avoidance methods that agencies can consider using and new techniques are continually being developed. These methods protect against disclosure in different ways and with different impacts on the resulting statistical products, making it challenging to compare and contrast the methods on an even footing.

To meet this challenge, this article proposes a scientifically principled way to examine the strengths and weaknesses of different disclosure avoidance methods in order to properly evaluate between them.

% These commands need to appear at the point where you want
% the first page to end.
\restoregeometry
\newgeometry{bottom=0.5in}

\section{Introduction}
\label{sec1}
The principal function of a statistical agency is to produce and release relevant, timely, credible, and objective statistical information (44 U.S.C. § 3563(a)(1)).\footnote{The principles laid out in this article could be applied to any organization faced with the need to protect the confidentiality of the data they steward. However, we will be focusing specifically on their application to a national statistical office, such as the U.S. Census Bureau.} In pursuing that mission, most statistical agencies have a countervailing obligation to protect the confidentiality of the data entrusted to them by their data subjects. While an agency can leverage many different statistical techniques and algorithmic frameworks to provide confidentiality protection, it is often unclear which approach is best for protecting a particular statistical product while maintaining the agency's objectives for disseminating the product. Without a clear set of criteria to use when comparing systems, selecting one approach over another can often be challenging and controversial. This article seeks to establish a set of objective principles that can inform statistical agencies' evaluation and selection of disclosure avoidance systems.

\section{What Is an Ideal, Applied Disclosure Avoidance System?}
\label{sec2}
A disclosure avoidance system (DAS) is a set of one or more statistical methods that transform confidential information from or about individual data subjects into statistics that describe the characteristics of groups without identifying the data subjects that comprise such groups. Disclosure avoidance systems accomplish this by applying statistical disclosure limitation (SDL) techniques to reduce (but not necessarily eliminate) disclosure risk in the statistical products being produced.

An \textit{applied} DAS is one that performs within the operational realities and production cycles of a statistical agency or unit\footnote{National statistical offices (NSOs) have a long history of designing and evaluating DAS, so much of the discussion in this article will center on the NSO perspective, but state and local agencies, nonprofit organizations, and private sector enterprises often face many similar challenges relating to data protection. To the extent that they do, the principles laid out in this article will also have relevance for these organizations.}. As such, it acknowledges and adapts to requirements stemming from the legal, policy, scientific, resource, and stakeholder environments it operates within. For example, an applied DAS may need to implement exogenously determined requirements such as statistical product design and schedule constraints.

An \textit{ideal}, applied DAS is one that both adapts to external requirements and conforms to a set of overarching principles or features relating to the efficiency, effectiveness, and flexibility of the system as it transforms confidential information from (or about) data subjects into quality statistics for public release. In most situations, such an \textit{ideal} DAS may not be fully achievable, but these principles can serve as a useful framework for evaluating the relative desirability and value of one candidate DAS over another.

\section{The Triple Tradeoff of Official Statistics}
\label{sec3}
At the heart of the challenge relating to disclosure avoidance is a set of interrelated objectives known as the “triple tradeoff of official statistics" \citep{hawes:2019, hawes:2021, abowd2023confidentiality, abowd2023sdl}, wherein agencies considering the release of a statistical product must balance between the competing dimensions of \textit{availability} of the statistics, \textit{accuracy} of those statistics, and \textit{confidentiality}. Every statistical product that an agency publishes that is derived from a confidential data source reveals or leaks confidential information in the process \citep{dinur:nissim:2003:10.1145/773153.773173}. Consequently, the more statistics an agency publishes (the \textit{availability}, granularity, and relevance of the statistics), and the greater the \textit{accuracy} of those statistics (the precision, validity, and reliability of the statistics), the higher the risk that they could permit \textit{confidentiality}-violating disclosure. Conversely, every statistical technique used to protect confidentiality degrades the resulting statistics in some manner. This is not a side effect: it is how SDL methods protect confidentiality. As a result, statistical agencies have to navigate an inherent tradeoff between the degree of confidentiality protection and the resulting availability and accuracy of the statistics to be released. When navigating this tradeoff, agencies must ask themselves questions such as `How much disclosure risk is too much?' or `How accurate, and at what level of granularity, do these statistics need to be to meet their intended (or legally mandated) uses?\footnote{Statistical agencies have long grappled with similar types of tradeoffs in the context of resource allocation for survey design and methodology: for example, `How much is the agency willing to spend on nonresponse follow-up to reduce coverage error?' When dealing with error from multiple sources, agency decisions for achieving necessary fitness-for-use targets will need to weigh the data protection dimension of the triple tradeoff in concert with other resource allocation decisions relating to other sources of survey error.}'

It is important to distinguish how the accuracy and availability dimensions of this triple tradeoff compare to related concepts, such as data \textit{usefulness} (or utility) and data \textit{quality}. Many discussions about the impact of SDL on statistical products eschew discussing data accuracy in favor of broader concepts like data usefulness in order to acknowledge that accuracy, as a traditionally defined concept describing the closeness of a statistic to its 'true' value, does not capture the myriad ways that the application of SDL may result in relatively 'accurate' statistics that otherwise fail to meet the needs of data users for specific use cases. For example, a poorly implemented suppression methodology can produce perfectly accurate and precise statistics (for those cells that are published), but with the lack of availability from such substantial and indiscriminate suppression of cells that would otherwise preclude valid statistical analysis. Similarly, a noise injection methodology may result in overall mean error of 0 while still yielding substantial variability or extreme outliers that could impede specific use cases. These discussions about accuracy become further complicated when one considers that many conventional discussions about the impact of SDL on data accuracy tend to only consider differences from the unprotected, confidential survey or census data, treating them as the `ground truth' while not considering the other sources of error (e.g., coverage error, measurement error) these statistical data collections already contain. The application of SDL may not have a substantial impact on the overall `accuracy' of the statistics as much as it results in deviations from the unprotected statistics. Those deviations may be large or small, but their relative impact on accuracy will depend on the magnitude of these other sources of error already present in the statistics. 

For these reasons, many authors tend to focus on the broader concept of utility, such as the `risk-utility framework' \citep{acdeb:2022, cnstat:2023}, `usability' \citep{hotz:2022}, or `fitness-for-use' \citep{abowd:2018}. In this regard, these terms are intended to describe how well the published statistics support their intended uses, particularly for those uses deemed to be of the highest value and importance to the nation. Adopting these broader conceptions of how 'useful' the resulting protected statistics are, however, obscures the complexity of the three-dimensional optimization that agencies perform under the triple tradeoff. Key aspects of this broader 'usefulness' concept relate to features of the statistics such as their granularity, coverage, and relevance: is the agency publishing statistics for critical (especially lower level) geographies such as places or school districts? Are variable categories being sufficiently disaggregated and are tables being sufficiently cross-tabulated to permit analysis of smaller demographic or socioeconomic groups? Is the agency publishing all of the tables that data users need to support their important use cases? These questions are certainly about the usefulness of the statistical product to meet the nation's needs, but they all actually pertain to the overall quantity (availability) of statistics being published. Holding the degree of confidentiality protection constant under the constraints of the triple tradeoff, any increase in the availability of statistics (to improve usefulness) necessitates decreased accuracy (reducing usefulness). 

The concept of data quality is likewise too broad to serve as an effective alternative for the accuracy dimension of the triple tradeoff. Data quality, as discussed in the Federal Committee on Statistical Methodology's (FCSM) Data Quality Framework, is "the degree to which data capture the desired information using appropriate methodology in a manner that sustains public trust" \citep{fcsm:2020}. As such, the concept of data quality encompasses more than just the accuracy and availability of the statistics, but also the confidentiality dimension of the triple tradeoff. Consequently, throughout the remainder of this article, we will use the terms \textit{accuracy} and \textit{availability} to describe the two respective dimensions of the triple tradeoff, we will use \textit{usefulness} in the general sense, recognizing that it encompasses both of those dimensions, and we will use the broader FCSM definition of data \textit{quality}, which further includes the confidentiality dimension. 

\section{Inherent Features versus Parameter Choices}
\label{sec4}
When evaluating the strengths and limitations of different DAS approaches, it is essential to distinguish the inherent characteristics of the systems from the implementation decisions and parameter choices made within those systems. Candidate DAS options should either be evaluated with comparable parameter choices, or they should be evaluated across a whole range of parameter settings, to ensure a fair `apples to apples' evaluation. For example, a DAS based on complementary cell suppression can offer a higher degree of data protection (with lower data availability) if it is implemented with a higher primary suppression threshold (e.g., suppress values under 10,000), or it can offer a lower degree of protection (with much higher data availability) if it is implemented with a lower primary suppression threshold (e.g., suppress values less than or equal to 3). Similarly, a rounding-based system can provide higher accuracy with less robust protections or lower accuracy with stronger protections, depending on whether it rounds values to the nearest 10 or 10,000. In this context, selecting a disclosure avoidance system based on suppression over one based on rounding should hinge on the characteristics and the relevant advantages and disadvantages inherent to those particular statistical disclosure limitation methods rather than on specific implementation or parameter choices for those systems under the triple tradeoff. Put another way, agency decision-makers can have a legitimate debate about the relative merits of suppression-based versus rounding-based disclosure avoidance systems, but it would be unfair for them to evaluate a suppression-based system with a primary suppression threshold of 10,000 against a rounding-based system that rounds to the nearest 10. Because many DAS approaches differ in substantial ways, selecting equivalent parameter comparisons may not always be feasible. After all, accuracy, availability, and confidentiality (degree of protection) are not simple, unitary concepts, and it is beyond the scope of this article to propose a unified metric that can equate every potential balance under the triple tradeoff with the corresponding set of parameter implementations for each differing SDL approach under consideration. In these cases, decision-makers should be cognizant of the impact that differences in parameters can make, and strive for equitable comparisons across DAS candidates.

The selection of both a DAS and parameters representing the agency's chosen balance under the triple tradeoff are crucial decisions with tangible consequences for the resulting quality of the statistical product to be released. An informed dialogue and debate about these choices should meaningfully distinguish between the system's characteristics and parameter selections. It is important to acknowledge, however, that no approach to disclosure avoidance will be perfectly adaptable to every potential system requirement; the selection of some methodologies over others may constrain the resulting systems' efficiency and flexibility to adapt to particular statistical product specifications, unique data user needs, or specific disclosure risks. Consequently, the evaluation of a DAS should not be viewed as a purely academic or theoretical exercise. While a principled evaluation of candidate systems requires distinguishing the inherent features of these systems from their subsequent implementation parameters, in practice, agency decision-makers will often approach these choices in a somewhat iterative fashion, adapting system design requirements, data product specifications, and implementation parameters as necessary to ensure the resulting quality of the statistical product.

\section{Principles of an Ideal, Applied Disclosure Avoidance System}
\label{sec5}
\subsection{Principle 1: It Should Support Meaningful Assessment of Disclosure Risk}
Central to the ability of a DAS to function is the ability to support meaningful assessment of disclosure risk. After all, how can an agency protect its data against unauthorized disclosure if it cannot objectively determine how likely a disclosure is to occur? 

In this context, however, the agency must be clear about what constitutes a disclosure and what does not. Many federal confidentiality statutes approach disclosure from the perspective of revealing information contained in a data subject's census or survey response (or in the context of administrative records or other third-party sources, information maintained by the statistical agency in the data subject's records). Section 9(a) of Title 13, for example, prohibits any data release ``whereby the data furnished by any particular establishment or individual under this title can be identified" \citep{title13}. Similarly, Section 3563 of the Foundations for Evidence-Based Policymaking Act of 2018 requires recognized statistical agencies and units to ``protect the trust of information providers by ensuring the confidentiality and exclusive statistical use of their responses" \citep{cipsea}. A simple reading of these requirements would seem to indicate that agencies would be prohibited from releasing any statistical products that would allow someone to determine a data subject's exact response to any particular census or survey question with complete certainty. 

Given the statutory frameworks, revealing a particular data subject's actual census or survey response would be an indisputable disclosure. Likewise, releasing a statistical product that permits deduction or inference about a data subject's response with perfect certainty would also be a clear disclosure, though here a statistical agency may need to begin making assumptions about the sophistication of data users or their access to external information. What about inference about a data subject's response with less than perfect certainty? If an agency merely modified one attribute on one data subject's record and then published the resulting data with the caveat that `one or more records have been modified,' would that trivial amount of uncertainty be sufficient to `protect' confidentiality under these statutes? If 99.99999\% certainty would constitute a disclosure, what about 99\% or 95\%? What about a coin toss (roughly 50\%)? Typically, there is no objective answer to this question. Rather, it falls into the realm of agencies' more subjective interpretations of their statutory responsibilities---and thus, their decision-making under the triple tradeoff. 

It is also important to note that not all inferences are confidentiality violating. Statistical agencies exist to produce statistical products that inform inferences about their respective societies. As such, the principal purpose of agencies' statistical products is to permit inference about those societies and their various groups and communities. Because membership in, and the characteristics of, those groups often follow statistical patterns, legitimate inference about those groups can inherently improve inference about a specific data subject's census or survey response. In the 2020 Census, for example, the racial and ethnic makeup of the state of Maine was 90\% non-Hispanic White. This simple statistic about the demographic environment of Maine would allow an attacker to infer with 90\% certainty that a randomly selected data subject within the state is non-Hispanic White. In assessing the impact of inference on disclosure risk, an ideal DAS must be able to differentiate inferences based on broader societal information that does not rely on the particular data subject's response from inferences informed by leakage from the data subject's record. Only the latter type of confidentiality-eroding inferences should be considered a disclosure risk, at whatever exogenously defined (i.e., agency determined) level of certainty.\footnote{It is outside the scope of this article to opine on how much disclosure risk should or should not be considered acceptable under the triple tradeoff, and that decision will likely differ depending on the agency's legal obligations, organizational risk tolerance, and the anticipated usefulness of the statistical product under consideration. It is similarly outside the scope of this article to describe how disclosure risk should be quantitatively defined and measured, though decisions about acceptable levels of risk certainly depend on those definitions and evaluation frameworks. For a thorough discussion of the principles a quantitative definition of disclosure risk should satisfy, see \citet{jarmin2023depth} and differing perspectives in \citet{hotz:2022}.} 

Having clarified what disclosure is (and is not) within the context of disclosure avoidance systems, it is now vital for a DAS to be able to assess the degree to which the uncertainty introduced by the DAS can mitigate that underlying disclosure risk. Otherwise, without the ability to assess disclosure risk and to measure uncertainty, there would be no way of knowing how much uncertainty would need to be introduced to achieve a desired level of risk mitigation. It should be noted, however, that the ability to assess disclosure risk and to measure uncertainty---both essential features of this principle---in no way prescribe or dictate how much risk mitigation is necessary or desirable. Put another way, the ability to assess disclosure risk and the corresponding level of uncertainty from SDL is a prerequisite for agencies to navigate the confidentiality/accuracy/availability dimensions of the triple tradeoff, but this feature of an ideal DAS does not determine the optimal point along any of those dimensions.

The underlying disclosure risk of a statistical product protected by a DAS can be assessed in different ways. Some systems can provide assessments of disclosure risk directly through analytical measures of the underlying uncertainty inherent to a statistical product's design and the additional uncertainty afforded by the application of SDL. Some coarsening SDL techniques, like uncontrolled, interior-cell rounding routines, can ensure that the most granular data disaggregations, from which all higher aggregations are built, limit precision to a specified degree of uncertainty. Similarly, properly implemented complementary suppression routines ensure that suppressed cells cannot be recalculated within a given degree of precision.\footnote{It should be noted, however, that disclosure risk assessments for suppression-based methods are brittle to external information or collusion between data subjects that can unravel the network of suppressions within a statistical product.} Formally private SDL approaches can also enable analytic assessments of uncertainty by measuring the underlying potential for leakage of response information in each released statistic. In each of these examples, the ability to assess uncertainty is independent of the agency's view on how much uncertainty is necessary or desirable. Rounding parameters, cell suppression thresholds, or formal privacy-loss budgets can be set high, low, or anywhere between. The challenge with these analytical approaches to uncertainty assessment, however, is that the uncertainty they measure does not necessarily directly correlate with tangible disclosure risk. Increasing uncertainty diminishes disclosure risk, but the ability of attackers to draw accurate inferences about data subjects' records in the presence of different levels and types of uncertainty also depends on the technology (both methodological and computational) and external information available to the attacker. Some mathematically possible attacks may not represent credible threats to an agency's disclosure avoidance protections because their underlying assumptions about computing power and external data may be unrealistic within reasonable time horizons. That said, ongoing advancements in technology might suggest that attempting to predict what types of attacks will or will not be feasible in the foreseeable future may be a dangerous form of speculation, particularly for more sensitive data collections.

An alternative method of assessing disclosure risk is through empirical assessment by simulating an actual attack on the statistical product's confidentiality protections and attempting to reveal or reidentify information about particular data subjects accurately. The most common form of these empirical assessments is reidentification studies on public-use microdata files or reconstruction-abetted reidentification studies on tabular products. However, increasingly sophisticated forms of simulated attacks (e.g., membership inference attacks) are also valuable techniques that should be considered. A major challenge with these empirical assessments is that because they are modeled on specific attack vectors, they are limited by the assumptions an agency makes about the assessed attack vector. Achieving low reidentification rates through one of these assessments does not necessarily mean that the underlying disclosure risk of a statistical product is actually low; it merely means that there is a low disclosure risk for the product against a particular attack vector under the prespecified assumptions and at the present moment. In spite of these limitations, however, empirical risk assessments remain a valuable tool for agencies to assess disclosure risk, because their results are often easy to interpret, and because they can confirm vulnerabilities that other assessment methods may merely suggest are possible, especially with regard to vectors of attack that agencies may consider especially worrisome \citep{JASON:2024, cnstat:2024}.

The current limitations of both empirical and analytical approaches to assessing disclosure risk mean that no existing disclosure avoidance system can perfectly achieve this principle. Despite these limitations, however, some systems are inherently better than others at achieving this principle's objectives. Properly applied controlled rounding mechanisms can better ensure baseline degrees of uncertainty than uncontrolled rounding mechanisms. Linear programming-based complementary suppression can provide guarantees that simple primary suppression algorithms cannot. Moreover, formally private noise-injection can provide provable measures of a potential attacker's uncertainty that other forms of noise injection cannot. 

\subsection{Principle 2: It Should Support Meaningful Assessment of the Impact of Data Protections on the Availability and Accuracy of the Statistics}
Under the triple tradeoff, applying any SDL technique to mitigate disclosure risk inevitably degrades the usefulness of the resulting statistical product along one or both of the other dimensions (availability or accuracy). Effective management of this tradeoff requires understanding and measuring what those impacts are. Otherwise, how can an agency decision-maker assess whether the resulting statistical product's usefulness to a nation is worth the release's marginal disclosure risk of publication? Impacts on the availability dimension of the tradeoff are often straightforward to measure---the quantity and granularity of the statistics being released can be easily observed. However, the impact of SDL on the accuracy of the resulting statistics can be much more difficult to effectively assess. 

Part of the difficulty of assessing the accuracy of statistics protected by a DAS stems from the fact that no single or uniform measure of \textit{accuracy}, broadly defined, exists. Abstractly, one might posit that the relative \textit{closeness} of the protected statistics to the unprotected data would be a good approximation of accuracy. But in the context of the uncertainty introduced through SDL (especially given other likely sources of error in the statistics), should that closeness be captured by measures of central tendency, the prevalence or magnitude of outliers, comparison relative to other known sources of error, or some other assessment of whether a consumer of the resulting statistics would reach a different conclusion in their analysis? 

Further complicating this assessment of post-SDL statistical accuracy is the fact that the degradation of accuracy caused by many SDL techniques (and the DASs that employ them) often depends on the underlying characteristics of the data to be protected. Examples of this data dependency can be seen across a wide variety of SDL techniques. Suppression-based disclosure avoidance techniques have greater degradation on statistical products that have more frequent cell counts below the suppression threshold. The impact of top- and bottom-coding methods (a common form of coarsening) largely depends on the prevalence of values above or below the recoding thresholds. Noise injection techniques that incorporate postprocessing for error reduction or enforced consistency can also have complex or surprising effects on accuracy depending on the underlying characteristics of the data being protected.\footnote{The Census Bureau's recent iterative efforts tuning the 2020 Census Disclosure Avoidance System's TopDown Algorithm (to assess and mitigate the system's potential to introduce bias in small counts and to evaluate the anticipated usefulness of the data for a wide range of stakeholder use cases) are an excellent example of the potential impact of this data dependency on statistical accuracy.\citep{uscb:ddps}} 

To be fair, disclosure avoidance systems that have data-dependent impacts on accuracy can still permit comprehensive evaluation of the accuracy of the resulting statistics---that assessment would merely need to be performed \textit{after} the protections have been applied. Reliance on ex post analysis of accuracy, however, is inefficient within the framework of the triple tradeoff. For any defined set of output statistics, agencies need to balance the degree of protection against the usefulness of the resulting statistics. If the impact of the SDL mechanism (and thus the resulting usefulness of the statistical product) is itself data dependent and cannot be effectively predicted in advance of the application of SDL, then the selected DAS may be an inefficient mechanism for implementing the chosen balance. The agency would either have to rely on an estimate of the impact (which represents an inherent imprecision in implementing the selected balance) or need to assess accuracy on multiple applications of the system at various levels and configurations of protection (which may be resource inefficient and could potentially increase the disclosure risk of the released product).

Ideally, agency decision-makers should be able to assess the full impact of disclosure avoidance on accuracy as part of their balancing of these dimensions under the triple tradeoff. The more precisely a DAS can anticipate the impacts on availability and accuracy before the application of SDL, the more effective and efficient that system will be at navigating the tradeoff.

\subsection{Principle 3: It Should Be Able to Target Protection}
One of the unfortunate realities of disclosure limitation, particularly in the demographic and economic contexts, is that disclosure risk is not evenly distributed across the underlying population. The more a particular data subject's attributes or characteristics differ from those around them, the easier it often is to single them out or reidentify them in a statistical product. Thus, for example, \textit{population uniques} have often been the primary focus of many DASs \citep{fcsm:dpt}.

Because of the increased vulnerability for these groups, an ideal DAS should be able to target increased protection where decision-makers deem it is most needed. A cynical interpretation of this principle might ask why some groups or individual data subjects should deserve greater protection than others, but the reality is that disclosure risk is not evenly distributed across society. A uniform approach to disclosure avoidance that achieves any arbitrary (but uniform) level of disclosure protection inherently overprotects some records, with the corresponding degradation of statistical product accuracy or availability under the triple tradeoff. Thus, equitable confidentiality protection can produce structured inequalities in the results \citep{donoharm}. Finding an optimal balance along those three dimensions of the triple tradeoff means being strategic and ethically minded about when and how more focused protections should be applied and about how strong (or weak) those targeted protections need to be. 

\subsection{Principle 4: It Should Be Able to Target Data Accuracy}
To give decision-makers the most flexibility in navigating the triple tradeoff, an ideal DAS should be able to achieve a range of protection from no protection to full protection, with corresponding data accuracy ranging from perfect accuracy to no accuracy. Achieving a proper balance along those dimensions should be an implementation decision, not a feature of the system. That said, the actual degradation in accuracy from any given degree of protection (the underlying assessment of which is covered by Principle 2) need not be distributed uniformly across all published statistics. Efficiently balancing accuracy and availability against the disclosure risk of a statistical product under the triple tradeoff relies mainly on the presumed importance of the statistical product to be released. That importance accrues from the use of those statistics. In situations where a single statistical product has many uses, it is often the case that some of those uses represent greater importance to the nation than others. Similarly, the relative impact of data protections on availability or accuracy may vary depending on the particular use case. That is to say, some uses of a statistical product may be more (or less) sensitive to the impacts of SDL-induced uncertainty on accuracy or coverage. Efficient balancing of these dimensions under the triple tradeoff means prioritizing the availability and accuracy for those uses that represent the highest societal importance, while recognizing that this may require greater degradation of accuracy in support of other lower importance uses in order to achieve the same overall level of data protection. An ideal, applied DAS should give decision-makers the most flexibility to achieve and implement the agreed upon balance that reflects those priorities. To that end, much like an ideal DAS allows decision-makers to target data protections for those most at risk of disclosure (rather than applying uniform protections that likely overprotect some records while underprotecting others), so too should the DAS allow decision-makers to target accuracy for those data uses with the highest importance to the nation.

\subsection{Principle 5: It Should Track Cumulative Disclosure Risk Over Time}
As noted above, every statistic that an agency releases that is derived from a confidential source carries a nonzero disclosure risk. While the risk from any single released statistic, especially at higher levels of aggregation, is often negligible, the cumulative risk across large volumes of data releases can be substantial. To effectively assess overall disclosure risk, it is important to assess the cumulative disclosure risk over these independent statistics as well as the potential interactions between them. Unfortunately, it is often these interactions between published statistics that complicate the cumulative assessment of disclosure risk. For example, two data tables protected using complementary cell suppression may each represent very low risks of disclosure on their own, but if a cell in one table can be used to infer the value of a suppressed cell in the other table, then the full set of primary and complementary suppressions can unravel through simple addition and subtraction. In this manner, the cumulative risk across both tables can be far greater than the sum of the individual disclosure risk of each table on its own. A DAS should be designed such that it is possible to assess the cumulative impact of each additional statistic, table, or product, while accounting for the potential interactions of statistics (and protections) across each successive release.

\subsection{Principle 6: It Should Be Transparent}
Statistical agencies have a professional and ethical obligation to be transparent about the statistical methods, techniques, and processes they employ and about any known or suspected limitations of the statistics they produce \citep{asa:2022, uscb:quality_standards}. Because any application of disclosure avoidance has unavoidable consequences for the resulting usability of data, an ideal, applied DAS should provide meaningful transparency to those who would use the resulting statistics. The DAS should be explainable to data users (especially those users without formal training in disclosure avoidance), to data subjects and the civil society groups that represent them, and to those who make and interpret the statutory frameworks within which the agencies operate. This includes providing clear information about the strength and nature of the confidentiality protections sufficient to inform data subjects about how their information is safeguarded. Conversely, this transparency should also extend to explanation of the impact of disclosure avoidance on the resulting statistics. Ideally, this would include quantitative assessment of disclosure avoidance--related uncertainty on those statistics and the provision of tools or methods to enable data users to factor this uncertainty into their statistical analyses. The system should also permit external assessment of the effectiveness of the confidentiality protections and of the correctness and efficiency of the system's implementation.

\subsection{Principle 7: It Should Be Feasible}
Statistical agencies operate within an environment of time-delimited research and production schedules and often of significant budgetary and human resource constraints. A DAS needs to be able to effectively and efficiently operate within these constraints. Therefore, an ideal, applied system should support deployment, customization, integration, testing, and operation within reasonable time frames and with reasonable expenditure of resources.

\section{Governance, Decision-making, and Parameter Selection}
\label{sec6}
These principles are intended to be a framework for agencies to evaluate potential DAS options. However, they are not intended to prescribe the use of one particular approach or system over another. There is often tension between these principles. For example, the sophistication of an approach that can improve risk assessment and meet specific accuracy targets can also make transparency or feasibility more challenging. Simple rounding rules are intuitive and easy to explain to nontechnical audiences, but are often of limited effectiveness when used to protect confidentiality for complex statistical products. Formally private methods may provide more effective and targeted disclosure protections, but their operation is often more difficult to explain.

Similarly, not all of these principles are necessarily equally important for an agency considering the release of a particular statistical product. For example, an agency may choose to place a higher value on transparency (Principle 6) compared to targeted protections for vulnerable data subjects or groups (Principle 3) because they may consider transparency to be more critical for the organization at that particular moment. Another agency may decide that assessing the impact of disclosure avoidance on accuracy (Principle 2) and the ability to achieve accuracy targets (Principle 4) are most critical in the context of a particular statistical product, and may place lower value on the remaining principles as a result.

Above all, these principles are intended to help differentiate decision-making about selecting a system (the particular SDL technique and algorithmic framework) from the implementation choices about the desired balance of the triple tradeoff.

It should also be noted that throughout this article there have been repeated references to `agency decisions' and `agency decision-makers' without clarifying how these decisions are being made or who the agency officials are making those decisions.

Technical expertise in statistical disclosure limitation is an essential component of agency decision-making under the triple tradeoff, but considerations regarding the availability and accuracy of statistical products are also critically important to this decision-making. Agency governance of decision-making regarding the selection of a disclosure avoidance system and decision-making regarding the desired balance under the triple tradeoff must, therefore, also reflect expertise in how the statistical products are intended to be used and the expected benefit to the nation to be gained from those uses. Consequently, the selection and implementation of a DAS should not be viewed as the exclusive purview of technical SDL practitioners; rather, these decisions should reflect the agency's broader priorities and collective expertise along all three of the dimensions of the triple tradeoff. Finally, while the responsibility and authority for finding and implementing an appropriate balance under the triple tradeoff ultimately rests solely with the agency itself, those who have to make these decisions will typically benefit from hearing the perspectives and viewpoints of the agency's internal and external stakeholders.

\section{The Critical Importance of Stakeholder Engagement}
\label{sec7}
Throughout this article, we have highlighted the foundational objective of statistical agencies to produce quality statistics that meet the nation's statistical needs, while also maintaining the trust of the data subjects that are reflected in those statistics by properly safeguarding the confidentiality of their information. As noted above, achieving this objective requires that agencies balance against the three competing dimensions of the triple tradeoff, weighing potential balances within this tradeoff against the importance of the statistical product that would be produced. But who determines what is most useful to the nation? In the most foundational sense, constitutional mandates, authorizing statutes, and legislative requirements provide clear, if incomplete, direction. Especially in the context of limited resources and competing priorities, agency decision-makers must typically identify what actions and products would provide the greatest benefit to the nation and guide the agency's activities accordingly. But agency decision-makers do not, and should not, make those decisions in a vacuum. Their thinking should be informed by the ongoing input and perspectives of a diverse landscape of stakeholders. These groups include those that directly or indirectly rely on the agency's statistical products, like federal, state, or local policymakers, researchers, businesses, and public health professionals, to name a few. They also include those who directly or indirectly support the agency's activities, including data subjects, appropriators, oversight bodies, civil society advocacy groups (including privacy advocates), and get-out-the-count organizations. All of these groups, and others, have their own valuable perspectives on the nation's statistical needs. These perspectives offer differing and contradictory implications for what statistical products should be developed, using which DAS, and with which implementation settings reflecting what specific balance under the triple tradeoff. None of these perspectives, including those of the agency decision-makers themselves, provides a complete picture of the overall value (and privacy costs) of a proposed statistical product. Only by regularly and actively engaging with these groups, discussing potential options and their implications, and evaluating the potential positive and negative consequences of agency decisions can those decision-makers hope to achieve an optimal balance under the triple tradeoff.

\section{Conclusion}
\label{sec8}
Selecting (or designing) a DAS is a challenging task. The choice of SDL techniques and the algorithmic framework for their deployment will typically have substantial consequences for the resulting system's ability to meet the statistical agency's objectives and requirements regarding the availability, accuracy, and confidentiality protection of its statistical products. The seven principles outlined above, which help identify the characteristics that such a system would ideally have, do not prescribe that any particular disclosure avoidance approach or system is inherently better or worse than another. Instead, they are intended to provide a framework for a statistical agency to use when evaluating between candidate systems that will allow it to determine, based on the agency's current priorities and objectives, which system will best assist the agency in producing and releasing high-quality statistical products---products that meet their data users' needs while also protecting the confidentiality of their data subjects.

\subsection*{Disclosure Statement}
The authors have no conflicts of interest to declare. Any opinions or viewpoints are the authors' own and do not reflect the opinions or viewpoints of the U.S. Census Bureau.

\subsection*{Acknowledgments}
The authors would like to thank Ron Jarmin, Christa Jones, danah boyd, Meghan Maury, Xiao-Li Meng, Ruobin Gong, Joseph Hotz, Ian Schmutte, the participants of the May 2024 National Bureau of Economic Research conference on Data Privacy Protection and the Conduct of Applied Research: Methods, Approaches and their Consequences, two anonymous reviewers, and our many colleagues at the U.S. Census Bureau, who provided invaluable feedback on earlier versions of this article.

\subsection*{copyright}No rights reserved. This work was authored as part of the Contributor’s official duties as an Employee of the United States Government and is therefore a work of the United States Government. In accordance with 17 U.S.C. 105, no copyright protection is available for such works under U.S. law. 

% All references should be stored in the file "references.bib"
% Please do not modify anything below this line.
\printbibliography

\end{document}